\documentstyle[12pt]{article}

\def\GeV{\,{\rm GeV}}

\def\sec{\,{\rm sec}}

\def\Mpc{\,{\rm Mpc}}

\def\cmm2{{\,\rm cm^{-2}}}
\def\cm2{{\,{\rm cm}^2}}
\def\cmm3{{\,{\rm cm}^{-3}}}
\def\gcmm3{{\,{\rm g\,cm^{-3}}}}

\def\mpl{{m_{\rm Pl}}}

\def\la{\mathrel{\mathpalette\fun <}}
\def\ga{\mathrel{\mathpalette\fun >}}
\def\fun#1#2{\lower3.6pt\vbox{\baselineskip0pt\lineskip.9pt
  \ialign{$\mathsurround=0pt#1\hfil##\hfil$\crcr#2\crcr\sim\crcr}}}

\begin{document}
\baselineskip=16pt
\pagestyle{empty}
\begin{center}
\rightline{FERMILAB--Pub--93/182-A}
\rightline{astro-ph/9307***}
\rightline{submitted to {\it Physical Review D}}
\vspace{.2in}

{\bf RECOVERING THE INFLATIONARY POTENTIAL}\\
\vspace{.2in}

Michael S. Turner \\
\vspace{0.1in}

{\it Departments of Physics and of Astronomy \& Astrophysics\\
Enrico Fermi Institute\\
The University of Chicago, Chicago, IL  60637-1433}\\
\medskip
{\it NASA/Fermilab Astrophysics Center\\
Fermi National Accelerator Laboratory, Batavia, IL  60510-0500}\\

\end{center}
\vspace{.3in}

\centerline{\bf ABSTRACT}
\bigskip

\noindent A procedure is developed for the recovery of the
inflationary potential over the interval
that affects astrophysical scales
($\approx 1\Mpc - 10^4\Mpc$).  The amplitudes of the scalar
and tensor metric perturbations
and their power-spectrum indices, which can in principle
be inferred from large-angle CBR anisotropy experiments
and other cosmological data,
determine the value of the inflationary
potential and its first two derivatives.  From these,
the inflationary potential can be reconstructed in
a Taylor series and the
consistency of the inflationary hypothesis tested.  A number of
examples are presented, and the effect of observational
uncertainties is discussed.

\newpage
\pagestyle{plain}
\setcounter{page}{1}

\section{Introduction}

The detection of anisotropy in the cosmic background radiation
(CBR) by the Differential Microwave Radiometer on the Cosmic
Background Explorer (COBE) satellite \cite{DMR} has provided the first
evidence for the existence of the primeval density perturbations
that seeded all the structure seen in the Universe today.  Two
other experiments have now confirmed the COBE detections \cite{confirm},
and numerous experiments are underway to probe anisotropy on angular
scales from arcminutes to tens of degrees.  [CBR anisotropy on
angular-scale $\theta$ arises primarily due to metric
perturbations on length-scale $100\Mpc\,(\theta /{\rm deg})$,
so that CBR anisotropy can probe metric fluctuations
on scales from about $10\Mpc$ to $10^4\Mpc$.]

The COBE DMR detection has opened the door for the study of the
primeval density perturbations, and thereby the microphysics
that produced them.  At the moment there are three viable models of
structure formation:  the cold dark matter models, wherein
the perturbations arise from quantum fluctuations
excited during inflation and expanded to astrophysical
length scales ($t\sim 10^{-34}\sec )$;
the models wherein the seed perturbations are
topological defects \cite{top}, such as textures,
cosmic strings, and global monopoles,
produced in a very-early phase transition ($t\sim 10^{-36}\sec )$;
and the PIB model \cite{pib}, wherein
the perturbations are local fluctuations in the baryon number
of unknown origin.  The PIB model distinguishes itself from
the others in requiring no nonbaryonic dark matter
($\Omega_0=\Omega_B\sim 0.2$).

The cold dark matter models motivated by inflation have been relatively
successful, though not without shortcomings \cite{jpo}.  In these
models there are, in addition to density (scalar metric) perturbations,
gravity-wave (tensor metric) perturbations that also give rise
to CBR temperature anisotropy.
This is a curse and a
blessing:  CBR anisotropy cannot be assumed to reflect
the underlying density perturbations alone; on the other hand, if
the tensor and scalar contributions can be separated, much
can be learned about the underlying inflationary potential
\cite{others,turner}.

The separation of the contribution of scalar and tensor perturbations
to CBR anisotropy involves
exploiting their different dependencies upon angular scale and possibly
their contributions to the polarization of
the CBR anisotropy \cite{davisetal,crit}.  In
addition, since the scalar perturbations alone seed the
formation of structure, measurements of the distribution
of matter in the Universe derived from red-shift
surveys, peculiar-velocity measurements, and so on can be used
to determine their spectrum independently.

The concern of this paper is what can be learned about the inflationary
potential from the spectral indices and amplitudes
of the scalar and tensor metric perturbations.  This question
has also been addressed elsewhere \cite{others}; our approach
follows the formalism set up in Ref.~\cite{turner} which
is applicable to inflationary potentials that are relatively smooth over
the interval that determines
metric perturbations on astrophysical scales.  It is
not applicable to potentials with ``specially engineered features''
\cite{designer}.

\section{The Method}

We use four observables to characterize
the scalar and tensor metric perturbations:
their contributions to the
variance of the CBR quadrupole anisotropy, $S$ for scalar
and $T$ for tensor, and the power-law indices of their
fluctuation spectra, $n$ for scalar and $n_T$ for tensor.
(For scale-invariant perturbations $n=1$ and $n_T=0$.
The horizon-crossing amplitudes of density perturbations
vary with scale as $\lambda^{(1-n)/2}$ and of the gravity-wave
perturbations as $\lambda^{-n_T/2}$.)

In Ref.~\cite{turner} it was shown that these
quantities can be related to the value of the
inflationary potential, its steepness, and the change in
its steepness, evaluated around the epoch that the
scales of current astrophysical interest crossed outside
the horizon during inflation (about 50 e-folds before
the end of inflation):
\begin{eqnarray}
V_{50} & \equiv & V(\phi_{50}) ;\nonumber\\
x_{50} & \equiv & {\mpl V^\prime (\phi_{50})\over V_{50}};\nonumber\\
x_{50}^\prime & \equiv & x^\prime (\phi_{50}) =
          {\mpl V^{\prime\prime}(\phi_{50})\over V_{50}}
          -{x_{50}^2\over \mpl} ;
\end{eqnarray}
$\phi_{50}$ is the value of the scalar field that drives
inflation 50 e-folds before the end of inflation (or however many
e-folds before the end of inflation the astrophysically
relevant scales crossed outside the horizon), $\mpl=1.22\times
10^{19}\GeV$ is the Planck mass, and prime indicates derivative
with respect to $\phi$.

The formulae relating the observables $S$, $T$, $n$, and
$n_T$ and the properties of the inflationary potential are
\begin{eqnarray}\label{eq:forward}
S \equiv {5\langle |a_{2m}^S|^2\rangle \over 4\pi} & = &
2.22{V_{50}\over \mpl^4 x_{50}^2}\left(1 +1.1(n-1)+{7\over 6}
[n_T-(n-1)]\right) ;\nonumber \\
T \equiv {5\langle |a_{2m}^T|^2\rangle \over 4\pi} & = &
0.606{V_{50}\over\mpl^4} (1 +1.2n_T) ;\nonumber \\
n & = & 1 -{x_{50}^2\over 8\pi} + {\mpl x^\prime_{50}\over 4\pi};\nonumber \\
n_T & = & -{x_{50}^2\over 8\pi} ;\nonumber \\
{T\over S} & = & 0.28x_{50}^2 = - 7\,n_T;
\end{eqnarray}
where $S$ ($T$) is the contribution of scalar (tensor) perturbations
to the variance of the CBR quadrupole temperature
anisotropy and brackets indicate the
ensemble average \cite{twl}.  Since
the four observables can be expressed in terms of three
properties of the potential a consistency check exists \cite{turner}.

These formulae
have been computed to lowest order in the deviation from scale
invariance, i.e., ${\cal O}(n_T, n-1)$, and
only apply to smooth potentials.  Note too that $n_T$ must
be less than zero (more power on large scales), though
the scalar power-law index $n$ can be greater than 1.
{}From Eqs. (\ref{eq:forward}) one can solve for the potential
and its first two derivatives:
\begin{eqnarray}\label{eq:recover}
V_{50} & = & 1.65\,T\,(1-1.2n_T) ;\nonumber \\
V_{50}^\prime & = & 5.01 \sqrt{-n_T}\,(V_{50}/\mpl ) ;\nonumber \\
V_{50}^{\prime\prime} & = & 4\pi\left[ (n-1)-3n_T \right]\,(V_{50}/\mpl^2).
\end{eqnarray}

At present, the COBE DMR detection serves mainly to determine
the sum of the scalar and tensor contributions to the
quadrupole anisotropy:\footnote{The value for
the variance of the CBR quadrupole anisotropy derived from
the COBE DMR data depends slightly upon
the spectral index of the metric perturbations; see Refs.~\cite{DMR}.}
\begin{equation}
T+S = \left( {16\pm 4\,\mu{\rm K}\over 2.726\,{\rm K}}\right)^2
\simeq 3.4\times 10^{-11}.
\end{equation}
In Ref.~\cite{turner} detailed formulae for the tensor and scalar
contributions to the higher multipoles are given; very roughly, for
$l\ll 200$ and standard recombination,
\begin{eqnarray}
{l(l+1)\langle |a_{lm}^S|^2\rangle\over 4\pi} &
\sim & S\,(l/2)^{n-1};\nonumber \\
{l(l+1)\langle |a_{lm}^T|^2\rangle\over 4\pi} & \sim & T\,(l/2)^{n_T}.
\end{eqnarray}
Thus, in principle, a separation of the tensor and scalar
contributions to the individual multipole amplitudes determines
$n_T$ and $n-1$.  Since $n_T$ is
directly related to the ratio of the tensor to scalar
contributions of the quadrupole anisotropy, measurements
of $S$ and $T$ also determines $n_T$.

The recovery of the inflationary potential proceeds by
constructing its Taylor series:
\begin{equation}\label{eq:taylor}
V(\phi ) = V_{50} + (\phi -\phi_{50})\,V_{50}^\prime
+ (\phi -\phi_{50})^2\,V_{50}^{\prime\prime}/2! + \cdots;
\end{equation}
as before, $\phi_{50}$ is the value of the scalar field
50 e-folds before the end of inflation.  Measurements
of $T$, $S$, $n$, and $n_T$ only
determine the square of $V_{50}^\prime$, so the sign of $V_{50}^\prime$
cannot be determined; as a matter of
convention we always take it to be negative.
The sign of $V^\prime$ is not physical since it can
be changed by the field redefinition:  $\phi \rightarrow -\phi$.

Scalar and tensor metric perturbations
on the astrophysically relevant scales---say from the
scale of galaxies, about $1\Mpc$, to the present horizon scale,
$H_0^{-1}\sim 10^4\Mpc$---were created during a small portion
of the inflationary epoch, corresponding to an interval
of roughly 8 e-folds around 50 e-folds before the end
of inflation (a precise formula relating the epoch when
a scale went outside the horizon during inflation
and the parameters of inflation
is given in Ref.~\cite{turner}).  This means that astrophysical
and cosmological data can only reveal information about the
inflationary potential over this narrow interval, a fact
which motivated the formalism developed in \cite{turner}.
As a matter of principle, we will only reconstruct
the potential over the interval that corresponds to these
8 e-foldings of the scale factor.

The equation of motion for $\phi$ in the slow-rollover
approximation \cite{slow}, $\dot\phi = -V^\prime /3H$, can be recast as
\begin{equation}
{d\phi\over dN} = {\mpl x\over 8\pi};
\end{equation}
where $N$ is the number of e-folds before the end of inflation.
By expanding the steepness $x$ around $\phi_{50}$,
$x(\phi ) = x_{50} + (\phi -\phi_{50})x_{50}^\prime$,
one obtains $\phi$ as a function of $N$:
\begin{eqnarray}\label{eq:phin}
\phi -\phi_{50} & = & {x_{50}\over x_{50}^\prime}\,
\left( \exp [(N-50)\mpl x_{50}^\prime /8\pi ] -1 \right) ; \nonumber \\
& = & {\mpl \sqrt{-n_T/2\pi} \over n-1 -n_T }
\left( \exp [(N-50)(n-1-n_T)/2]-1 \right).
\end{eqnarray}
The change in the value of the scalar field over the
8 important e-folds of inflation ($=\Delta\phi$) depends upon $n$ and
$n_T$:  If the difference between $n-1$ and $n_T$ is very
small, then $\Delta \phi \sim \sqrt{-n_T}\mpl$; on the other
hand, if $\sqrt{-n_T}$ is very small or the difference
between $n-1$ and $n_T$ is large, then $\Delta\phi$ is much less than $\mpl$.

This equation, together with the Taylor expansion
for the potential, cf. Eq.~(\ref{eq:taylor}), and
the equations relating $V_{50}$, $V_{50}^\prime$,
and $V_{50}^{\prime\prime}$ to
the observables $S$, $T$, $n$, and $n_T$, cf. Eqs.~(\ref{eq:recover}),
are all we need to recover the inflationary potential.
To begin, we will recover some familiar inflationary potentials which
have been analyzed elsewhere in the formalism discussed
above \cite{turner}.  For these potentials
we do not worry about the scale of inflation, $V_{50}$,
which is set measurements of $S$ and $T$ (see below); we
will only be interested in the shape of the potential.
Specifying $n_T$ and $n$ is sufficient to recover
the shape, though we also give $T/S$ as
it may be easier to measure than $n_T$ (and of course
is equivalent to $n_T$).

\section{Some Examples}
\subsection{Familiar potentials}
First, consider potentials of the form $V(\phi ) = a\phi^b$,
often used in models of chaotic inflation \cite{chaotic}.
For these models \cite{turner}
$$T/S=0.07b;\qquad n_T = -0.01b; \qquad n = 0.98 - 0.01b. $$
Note, the deviations from scale invariance increase with $b$;
since our recovery process involves an expansion in the
deviation from scale invariance one expects the recovery
of the potential to
be less accurate for larger values of $b$.  In Figs. 1
we show the original potential and the recovered potential for
$b =2,4,16$; even for $b=16$ the recovery is quite accurate.

Next, consider exponential potentials, $V(\phi ) =
V_0 \exp (-\beta\phi/\mpl )$, which arise in models of
extended inflation \cite{extended}.  For these models \cite{turner}
$$T/S=0.28\beta^2;\qquad n_T = -{\beta^2\over 8\pi};
\qquad n-1 =n_T.$$
In Figs. 2 we show the reconstruction for $\beta
= 1.23, 1.94, 6.03$, corresponding to
$n_T= -0.06, -0.15, -0.24$.  Only for
$n_T =-0.24$ is the recovery of the potential less excellent;
however, this much deviation from scale
invariance is probably inconsistent with
models of structure formation \cite{tilt}.

Now, consider a cosine potential, $V(\phi )=
\Lambda^4 [1+\cos (\phi /f)]$, the type of potential
employed in the ``natural-inflation'' models \cite{natural}.
It is not possible to provide a general formula for
$n_T$, $n$, and $T/S$; however, there are two limiting regimes:
$f\ga \mpl$ and $f\la \mpl$.  In the first regime, the
cosine potential reduces to the case of chaotic inflation
with $b=2$.  In the second regime \cite{turner},
$${T\over S} = 0.07\left({\mpl \over f}\right)^2
\left({\phi_{50}\over f}\right)^2\ll 1; \qquad n =1 -{1\over 8\pi}
\left({\mpl \over f}\right)^2;$$
where $\phi_{50}/f\simeq \pi \exp (-50\mpl^2/16\pi f^2)$.
In Fig.~3 we show the recovered potential for $f=\mpl /2$,
where $n=0.84$.  Again, the recovery process works very well.

Finally, consider the Coleman-Weinberg potentials,
$V(\phi )= B\sigma^4/2+B\phi^4[\ln (\phi^2/\sigma^2)-0.5]$,
often used in models of new inflation; for these models
$${T\over S} \simeq 3\times 10^{-5}\left( {\sigma \over
\mpl}\right)^4;\qquad n=0.94 .$$
These potentials are extremely flat and easily recovered
as shown in Fig.~4.
\subsection{Unknown potentials}
Now we turn to the recovery of an unknown potential
from cosmological data.  The recovery process requires knowledge of
three of the quantities $T$, $S$, $n$, and $n_T$;
we will use $T+S$, $T/S$, and $n$, which are probably the easiest to measure.
The quadrupole temperature anisotropy measures $S+T$; supposing that
its value is $16\,\mu{\rm K}$, the COBE DMR determination,
we can immediately infer $V_{50}$:
\begin{equation}\label{eq:v50}
V_{50} = (3.3\times 10^{16}\GeV )^4\,
\left( { 1 -1.1(n-1)+{7\over 6}
(n-1-n_T ) \over 1 + S/T }\right) .
\end{equation}
(We remind the reader that $n_T = -0.14T/S$.)
{}From this equation we see that the value of $V_{50}$ is
most sensitive to $T/S$, varying inversely with it.  That
is, the scale of inflation rises with the amplitude of
tensor perturbations, asymptotically approaching an
energy scale of about $3\times 10^{16}\GeV$.

Once $V_{50}$ is fixed, $n$ and $T/S$
determine the shape of the potential.
Generically, there are four qualitatively different
outcomes for the measured quantities which lead to
four generic inflationary potentials:
\begin{enumerate}

\item $n \approx 1$ and $T/S$ very small, corresponding
to scale-invariant scalar and tensor perturbations

\item $n$ significantly less than 1 and $T/S$ very small, corresponding
to tilted scalar fluctuations and scale-invariant, small-amplitude
gravity waves

\item $n\approx 1$ and $T/S$ of order unity,
corresponding to scale-invariant scalar perturbations and
tilted, large-amplitude gravity waves

\item $n$ significantly less than 1 and $T/S$ of order
unity, corresponding to tilted scalar and tensor perturbations
and large-amplitude gravity waves
\end{enumerate}

The four generic potentials are illustrated in Figs.~5.  For
large $T/S$, cases (3) and (4), the potential is
steep, the scale of inflation is relatively large, and
the variation of $\phi$ over the relevant 8-folds is
of the order of the Planck mass.  For small $T/S$, cases
(1) and (2), the potential is very flat, the
scale of inflation is relatively low, and the variation
of $\phi$ over the relevant 8-folds is much less than the Planck mass.
Coleman-Weinberg potentials provide an example of case (1);
cosine potentials and the potential $V(\phi )
= -m^2\phi^2+\lambda\phi^4$ \cite{slow}
provide examples of case (2); recently,
an example of a potential corresponding to case (3) has
been presented \cite{barrow}; and exponential
potentials provide an example of case (4).  Finally,
$n$ can be larger than unity;
however, the two new cases, $n$ significantly greater than one
and $T/S$ small or of order unity,
are qualitatively similar to cases (2) and (4).

\section{Discussion}

The scalar and tensor contributions to the CBR quadrupole
anisotropy, $S$ and $T$,  and the power-law indices of the spectra of
scalar and tensor perturbations, $n$ and $n_T$, serve to determine---indeed
overdetermine---the value of the inflationary potential
and its first two derivatives.  In principle,
measurements of these four quantities
can be used both to test the consistency of the
inflationary hypothesis and to recover the inflationary
potential through the first three terms in its Taylor expansion.
We have shown the recovery of several familiar potentials,
cf. Figs.~1-4, and the four generic types of inflationary
potentials that arise, cf.~Figs.~5.

In order to recover the inflationary potential
measurements of at least three
of the quantities $n$, $T$, $S$, and $n_T$ are required.
In all likelihood the first three will be the easiest to
determine; CBR anisotropy as well as determinations
of the distribution of matter and large-scale structure
should serve to measure $n$, and large-angle CBR anisotropy
should determine $S$ and $T$
(in the case of $T$, at least an upper limit).  An independent
measurement of $n_T$ seems much more difficult,
but provides an important consistency check.

In any case, determinations of $n$, $S$, and $T$ are likely to
have significant uncertainties, so that the recovery of the
underlying inflationary potential will not be as easy
or precise as our examples would indicate.  In Fig.~6, we show the effect
of these uncertainties on the recovery of the shape
of the inflationary potential for the following data:
$n=0.9 \pm 0.2$ and $T/S =0.3\pm 0.25$.  Even worse is the
effect of uncertainties on determining the scale of
the potential:  Recall, when $S+T$ is normalized
to the COBE result, $V_{50}$ varies
as the inverse of $T/S$, which for the above ``data'' leads
to an order of magnitude range in the value of $V_{50}$.

An accurate recovery of the inflationary potential is still
a long way from reality---and, of course, it may be that inflation
never even occurred.  However, with the COBE DMR anisotropy measurements
the first step has been taken.  Moreover, the potential payoff---probing
physics at unification energy scales---is worth the effort, if not
the wait.

\vskip 1.5cm
\noindent  This work was supported in part by the
DOE (at Chicago and Fermilab) and by the NASA through
NAGW-2381 (at Fermilab).  This work was completed at
the Aspen Center for Physics.

\newpage

\begin{center}
{\bf FIGURE CAPTIONS}
\end{center}
\medskip

\noindent {\bf Figure 1:}  Recovery of chaotic potentials,
$V(\phi ) =a\phi^b$, over the 8 e-folds relevant
for astrophysical scales and for comparison the original potential
(broken curves):  (a) $b=2$; (b) $b=4$; and (c) $b=16$.

\bigskip
\noindent{\bf Figure 2:}  Recovery of exponential potentials,
$V(\phi )=V_0 \exp (-\beta\phi /\mpl )$,
and for comparison the original potential (broken curves):
(a) $\beta =1.23$; (b) $\beta =1.94$; and (c) $\beta = 6.03$.

\bigskip
\noindent{\bf Figure 3:}  Recovery of the cosine potential,
$V(\phi )=\Lambda^4[1+\cos (\phi /f)]$ and $f=\mpl /2$,
and for comparison the original potential (broken curve).

\bigskip
\noindent{\bf Figure 4:}  Recovery of a Coleman-Weinberg
potential with $\sigma =1\times 10^{16}\GeV$ and for comparison
the original potential (broken curve).

\bigskip
\noindent{\bf Figure 5:}  The four generic inflationary
potentials:  (a) $n-1=-2\times 10^{-6}$ and $T/S=1.4\times 10^{-5}$, with
the COBE DMR normalization $V_{50}^{1/4}= 2.0\times 10^{15}\GeV$;
(b) $n=0.85$ and $T/S=1.4\times 10^{-4}$, $V_{50}^{1/4}=
3.6\times 10^{15}\GeV$; (c) $n=1$ and $T/S=1$, $V_{50}^{1/4}=
2.9\times 10^{16}\GeV$; and (d) $n=0.85$ and $T/S=1$,
$V_{50}^{1/4}=2.9\times 10^{16}\GeV$.

\bigskip
\noindent{\bf Figure 6:}  An illustration of
the effect of observational uncertainties on the shape of
the recovered potential; here $n=0.9\pm 0.2$ and $T/S =0.3\pm 0.25$.
The four curves correspond to $n=0.7$ and $T/S=0.05$ (solid),
$n=0.7$ and $T/S=0.55$ (dotted), $n=1.1$ and $T/S = 0.05$ (dashed), and
$n=1.1$ and $T/S=0.55$ (long-dashed).

\end{document}